\documentclass[aps,prl,twocolumn,amsmath,amssymb,superscriptaddress,10pt]{revtex4-2}
\usepackage{fix-cm}
\usepackage{array}
\usepackage{bm}
\usepackage{color}
\usepackage{float}
\usepackage{graphicx}
\usepackage[caption=false]{subfig}

\usepackage{newtxtext}
\usepackage{newtxmath}
\usepackage{braket}
\usepackage{enumitem}
\usepackage{gensymb}
\usepackage{natbib}
\usepackage{soul}
\usepackage{lipsum}
\usepackage{hyperref}
\hypersetup{colorlinks=true,allcolors=blue}

\begin{document}
\title{Quantum Metric Localization and Quantum Metric Protection }

\author{Wen-Bo Dai}\thanks{These authors contributed equally to this work.}
\affiliation{Department of Physics, Hong Kong University of Science and Technology, Clear Water Bay, Hong Kong, China}
\author{Jinchao Zhao}\thanks{These authors contributed equally to this work.}
\affiliation{Department of Physics, Hong Kong University of Science and Technology, Clear Water Bay, Hong Kong, China}
\author{Shuai A. Chen} \thanks{These authors contributed equally to this work.}
\affiliation{Max Planck Institute for the Physics of Complex Systems, N\"{o}thnitzer Stra{\ss}e 38, Dresden 01187, Germany}
\author{K.T. Law}
\email{phlaw@ust.hk}
\affiliation{Department of Physics, Hong Kong University of Science and Technology, Clear Water Bay, Hong Kong, China}


\begin{abstract}
The study of disorder effects in electronic systems is one of the central themes in physics. It is well established that in the Anderson localization regime, the localization length of electrons decreases monotonically as the disorder strength increases. Here, we demonstrate that the conventional Anderson localization paradigm fails completely in describing an isolated band with quantum metric, where the quantum metric of the band defines a length scale called the quantum metric length. For an isolated band with a finite bandwidth separated from other bands by a band gap $\Delta$, weak disorder results in conventional Anderson localization behavior. However, as the disorder increases, the localization length ceases to decrease and becomes pinned at a value proportional to the quantum metric length, forming a localization length plateau.
We term the regime within this localization length plateau as the quantum metric localization regime. Remarkably, the localization length does not deviate from the plateau until the disorder strength far exceeds $\Delta$. We refer to this strong protection against disorder, characterized by the quantum metric length, as quantum metric protection. In this work, we first numerically demonstrate quantum metric localization using a 1D Lieb lattice. We then provide a simple physical picture based on the properties of Wannier functions to explain the origin of the localization length plateau. A supersymmetric field theory approach explains why the localization length is proportional to the quantum metric length and captures the crossover from Anderson localization to quantum metric localization. Our conclusions are broadly applicable to disordered electronic, photonic, and acoustic systems.
\end{abstract}
\date{\today}

\maketitle

\section*{Introduction}
The study of the quantum geometric properties of materials has emerged as a central theme in modern condensed matter physics. The quantum geometric properties of Bloch states are characterized by the quantum geometric tensor \cite{provost1980riemannian}, whose imaginary part corresponds to the Berry curvature and the real part defines the quantum metric. Over the past several decades, Berry curvature effects have been extensively explored and have proven to be of fundamental importance across a broad range of physical phenomena \cite{haldane2004berry,nagaosa2010anomalous,XiaoRMP2010, sodemann2015quantum, resta1994macroscopic, thonhauser2005orbital, essin2009magnetoelectric,qi2008topological,Vanderbilt_2018}. Remarkably, the integral of the Berry curvature over the filled bands of a two-dimensional system yields a quantized topological invariant—the Chern number—which characterizes the bulk properties of an insulating state \cite{thouless1982quantized}. Physical observables tied to a finite Chern number, such as the quantized anomalous Hall conductance \cite{changQAH2013, CCzRMP2023}, remain robust against perturbations as long as the bulk energy gap persists. In this sense, the Berry curvature gives rise to rigid topological protection.

\begin{figure*}[ht]
    \centering
    \includegraphics[width=17.2cm]{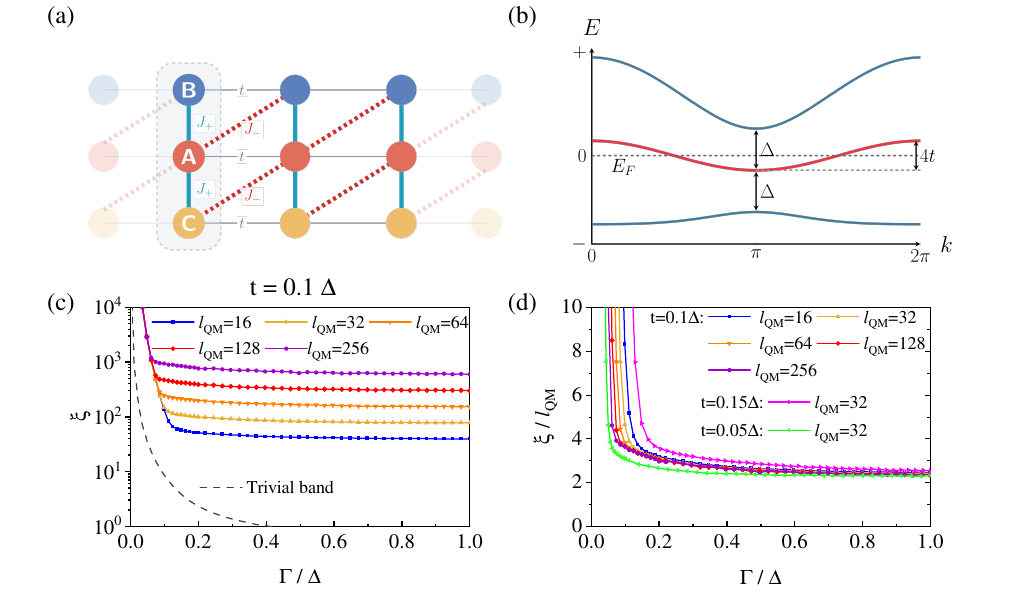}
    \caption{(Color online).   
\textbf{(a)} Schematic of the Lieb lattice, characterized by vertical and diagonal hopping amplitudes $J_+$ and $J_-$, together with a uniform horizontal hopping $t$ that governs the band structure of the central band. 
\textbf{(b)} Energy spectrum in the clean limit. The Fermi level $E_F$ (dashed line) lies within the central band, which serves as the focus for studying transport properties governed by the quantum metric.
\textbf{(c)} Disorder dependence of the localization length $\xi$, showing the localization length plateau for a wide range of disorder. The value to the plateau is determined by $l_{\mathrm{QM}}$. For comparison, for a trivial band (dashed line) without quantum metric, the localization length drops to the atomic limit when the disorder strength is comparable to the bandwidth $4t$. 
\textbf{(d)} Rescaling $\xi$  by $l_{\mathrm{QM}}$ yields a collapse of $\xi/l_{\mathrm{QM}} \sim 2$, forming a localization plateau across different dispersions and quantum metric lengths. The parameters are chosen to fix the band gap $\Delta = 2\sqrt{2}J\delta =20$ while quantum metric tuned via the aniosotropic parameter
$\delta$ on a chain with $10^7$ sites
}
    \label{fig:Lieb}
\end{figure*}

On the other hand, the profound physical significance of the real part of the quantum geometric tensor—the quantum metric—has come to be fully appreciated only relatively recently. The quantum metric, which encodes the gauge-invariant distance between neighboring quantum states in the parameter space, has been shown to govern a remarkably diverse set of phenomena \cite{marzari1997maximally,marzari2012maximally,verma2026quantum, anandan1990geometry,liu2025quantum,rossi2021quantum,yu2025quantum,gao2014field,wang2023quantum,gao2023quantum,raoux2015orbital,gao2015geometrical, ahn2022riemannian,kang2025measurements,topp2021light, roy2014band, antebi2024drude,mitscherling2022bound,bouzerar2022giant,kruchkov2023quantum}. The quantum metric effects in flat band materials with small Fermi velocity are particularly interesting. For example, it has been demonstrated that the quantum metric generates a finite superfluid stiffness \cite{peotta2015superfluidity,torma2022superconductivity,chen2024ginzburg} even in exactly flat bands where conventional kinetic contributions vanish entirely. Moreover, many electronic length scales associated with the Fermi velocity vanish in flat band materials according to conventional theories. Recently, it was shown that the superconducting coherence length \cite{chen2024ginzburg, hu2025anomalous}, the coherence length of Josephson junctions \cite{li2025flat}, and the localization length of Majorana modes \cite{guo2025majorana} are all governed by the quantum metric length. It is well-known that in 1D the Anderson localization length $\xi$ is proportional to $v_F^2/\Gamma^2$ where $v_F$ is the Fermi velocity and $\Gamma$ is the disorder strength \cite{anderson1958absence,thouless1974electrons,Thouless1977,abrahams1979scaling,Lee1985,Beenakker1997RMP}. An outstanding question is: What is the localization length in disordered flat band materials when $v_F$ is small or even vanishes?

Several pioneering works have contributed to the understanding of disordered flat band materials \cite{nishino2007flat,leykam2013flat,leykam2017localization,MitscherlingPRB2020,Bouzerar2020PRR,BouzerarPRB2021,vcadevz2021metal,MitscherlingPRB2022,MeraPRB2022,bouzerar2022giant,Onishi2025,RosenPRX2025,wang2026qml,HZPRB2026}, exploring concepts such as disorder-enabled diffusion \cite{wang2026qml,Burkov2026,yin2026classical_percolation_quantum_metric}. However, the nature of the localization and the value of the localization length in these systems remained elusive. In this work,we found a new localization regime which cannot be described by Anderson localization.  We demonstrate that for a wide range of disorder strength, the localization length stays constant and is governed by the quantum metric length $l_{\mathrm{QM}}$ of the flat band as defined in Eq.~\ref{eq:lqm}. 

Specifically, we begin by examining a 1D Lieb lattice, illustrated in Fig.~\ref{fig:Lieb}a. As its band structure shows, the nearly flat band at the Fermi energy $E_F$ possesses a bandwidth of $4t$ and is separated from other bands by an energy gap $\Delta$. We subsequently introduce on-site uniform disorder with strength $\Gamma$ to the Lieb lattice. As shown in Fig.~\ref{fig:Lieb}c, the localization length $\xi$ approaches a robust plateau as the disorder strength increases. The magnitude of this localization length plateau is governed by $l_{\mathrm{QM}}$. Crucially, when $\Gamma$ is comparable to the bandwidth $4t$, the localization length of a trivial band (zero quantum metric) rapidly collapses to the atomic limit (dashed line), as conventionally expected. In stark contrast, the localization length of a band with nontrivial quantum metric can be orders of magnitude longer and remains independent of disorder strength, even when $\Gamma$ approaches $\Delta$. Incredibly, as depicted in Fig.~\ref{fig:Lieb}d, when the localization length is rescaled by $l_{\mathrm{QM}}$, all curves corresponding to different values of $l_{\mathrm{QM}}$ and bandwidths collapse onto a single universal plateau at $\xi \sim 2l_{\mathrm{QM}}$.

The results shown in Fig.1 undoubtedly suggest that the quantum metric length is a fundamental length scale governing the localization properties of flat band materials. We refer to the regime where the localization length lies on the plateau governed by the quantum metric as the quantum metric localization regime. This is in sharp contrast to the Anderson localization regime \cite{anderson1958absence,abrahams1979scaling} where the localization length decreases monotonically as the disorder strength increases. In Fig.~\ref{fig:Lieb}d, we can clearly see the crossover from the Anderson localization regime to the quantum metric localization regime as disorder increases.

Furthermore, as shown in Fig.~\ref{fig:plateau_dissolve}, the localization length plateau is determined by the quantum metric length and is independent of bandwidth $4t$ and the band gap $\Delta$. The localization length deviates from the plateau only when the disorder is so strong that states from faraway bands contribute significantly to the formation of localized states at the Fermi energy. Therefore, the integral of the quantum metric over the Brillouin zone, the quantum metric length $l_{\mathrm{QM}}$, is a protected quantity characterizing the properties of the Bloch states of the isolated band. We call this protection of the quantum metric length against disorder the quantum metric protection. 
We further demonstrate the universality of this phenomenon in the Appendices by showing that quantum metric protection persists in a spinful model subject to magnetic impurities, and in a two-band model where the kinetic dispersion and geometric metric can be completely decoupled.

Several distinct features of the quantum metric protection merit emphasis. First, unlike topological protection, which relies on global topological invariants in gapped systems, quantum metric protection is governed by the gauge-invariant length scale defined by states in a sub-Hilbert space. This quantum metric protection remains effective even in topologically trivial or gapless systems. Second, topological protection is often associated with quantized physical quantities such as quantized Hall resistance or quantized tunneling conductance. The protected physical quantities in the quantum metric protection (such as $l_{\mathrm{QM}}$) are highly tunable. Third, when $l_{\mathrm{QM}}$ is comparable to system size, electrons evade localization even for strong disorder. Therefore, $l_{\mathrm{QM}}$ emerges as a crucial length scale for mesoscopic systems. Finally, because this quantum metric protection is fundamentally a wave phenomenon, our results and theoretical framework are broadly applicable to photonic waveguide arrays \cite{schwartz2007Nature,LahibiPRL2008}, acoustic resonator networks \cite{xueNM2019}, and superconducting Josephson junction networks \cite{RosenPRX2025}. 

The remainder of this paper is organized as follows. In Sec. II, we introduce the modified 1D Lieb lattice and numerically demonstrate the localization plateau and its universal scaling collapse. Section III provides an intuitive physical picture based on Wannier functions and Thouless formula, showing how the quantum metric enables disorder-induced diffusion. In Sec. IV, we develop a supersymmetric non-linear sigma model that analytically yields the plateau localization length $\xi \sim \mathcal{O}(1)l_{\mathrm{QM}}$ and captures the crossover from Anderson localization to quantum metric localization. The Appendices contain technical derivations for the field theory, including the treatment of finite dispersion and multiband effects, and further demonstrate the universality of our findings through an independent two-band model and a spinful Lieb lattice subject to magnetic disorder. We conclude in Sec. V with a discussion of quantum metric protection and its broader implications.

\section{Model and Numerical Scaling Collapse}
\subsection{The Modified Lieb Lattice and Quantum Geometry}
To demonstrate the emergence of a fundamental length scale—the quantum metric length—beyond the Anderson localization paradigm, we consider a modified one-dimensional Lieb lattice, as shown in Fig.~\ref{fig:Lieb}a. The lattice is composed of unit cells labeled by $x$, each hosting a three-component sublattice vector $\mathbf{c}_x = (c_{x,A}, c_{x,B}, c_{x,C})^T$, where the components correspond to the $A, B$, and $C$ sublattices of the Lieb lattice respectively. The lattice constant is set to $a=1$ throughout this work. The system is described by the tight-binding Hamiltonian
\begin{equation}\label{eq:Hlieb}
H =
\sum_{x}
\mathbf{c}_x^\dagger \hat{H}_0 \mathbf{c}_x
+
\sum_{x}
\left(
\mathbf{c}_{x}^\dagger \hat{H}_1 \mathbf{c}_{x+1} + \mathrm{h.c.}
\right)
+
\sum_{x}
\mathbf{c}_x^\dagger \hat{V}_x \mathbf{c}_x.
\end{equation}
Explicitly, the intra-cell and inter-cell hopping matrices read
\begin{equation}
\hat{H}_{0} =
\begin{pmatrix}
0 & J_+ & J_+ \\
J_+ & 0 & 0 \\
J_+ & 0 & 0
\end{pmatrix},
\qquad
\hat{H}_{1} =
\begin{pmatrix}
t & J_- & 0 \\
0 & t & 0 \\
J_- & 0 & t
\end{pmatrix},
\end{equation}
where $J_\pm = (1 \pm \delta)J$, with $J$ the uniform hopping amplitude and $\delta$ a dimerization parameter.

To understand the transport behavior under disorder, we incorporate on-site Anderson disorder into the lattice, represented by $\hat{V}_x=V(x)\mathbf{I}_{3\times3}$ in the real space orbital basis. Here, $V(x)$ is a random potential following a uniform distribution with zero mean and white-noise correlation $\langle V(x)V(x') \rangle_{\text{dis}} = \Gamma^2 \delta_{x,x'}$, where $\Gamma$ denotes the disorder strength.

In the clean limit ($\hat{V}_x = 0$), $H(k)$ yields a three-band spectrum: a central isolated band positioned around zero energy, bounded by two dispersive bands and separated by a bulk energy gap $\Delta= 2\sqrt{2}J\delta$ with corrections of order $\mathcal{O}(t)$ [see Fig.~\ref{fig:Lieb}b]. In the following studies, the Fermi energy $E_F$ is set within the central band, which possesses a small kinetic dispersion $E(k)=2t\cos(k)$ with a bandwidth of $4t$. The quantum geometry of a sub-Hilbert space corresponding to a band is characterized by the quantum metric
\begin{equation}
g_{xx}(k) = \langle \partial_k u_f | \partial_k u_f \rangle - |\langle u_f | \partial_k u_f \rangle|^2,
\end{equation}
where $|u_f\rangle$ is the corresponding Bloch state of the central band.
The quantum metric length ($l_{\mathrm{QM}}$) is defined as the Brillouin zone average of the local quantum metric $g_{xx}(k)$ such that
\begin{equation}\label{eq:lqm}
l_{\mathrm{QM}} \equiv \bar{g}_{xx} = \int \frac{dk}{2\pi} g_{xx}(k).
\end{equation}
This length scale is independent of the energy dispersion and encodes the fundamental quantum geometric properties of the Hilbert space of the isolated band. In our modified Lieb lattice model, it simplifies to: $l_{\mathrm{QM}}=\frac{(1-\delta)^2}{8\delta}$.

\begin{figure*}[t]
    \centering
    \includegraphics[width=17.2cm]{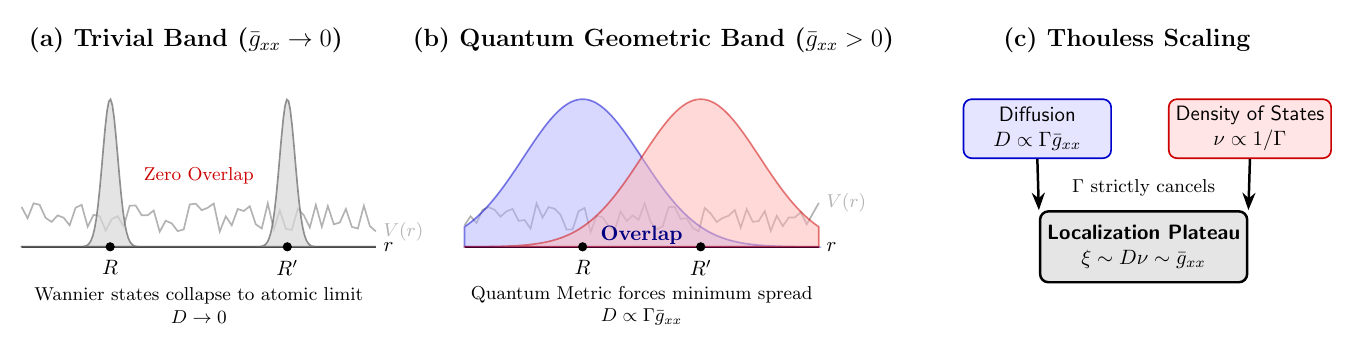}
    \caption{(Color online). Quantum metric enabled diffusion picture and localization length plateau. \textbf{(a)} In a trivial band ($\bar{g}_{xx}\rightarrow0$), optimally localized Wannier states can collapse to the atomic limit. The spatial density overlap between adjacent sites vanishes, shutting down disorder-induced hopping and diffusion ($D\rightarrow0$). \textbf{(b)} The nontrivial quantum metric ($\bar{g}_{xx}>0$) enforces a lower bound on the spatial spread of the Wannier states, guarantees a finite overlap between the Wannier states, driving a diffusion constant $D\propto\Gamma\bar{g}_{xx}$. In other words, finite quantum metric length is essential for diffusion. \textbf{(c)} The Thouless scaling bridging the diffusion to the localization length. The disorder scattering rate $\Gamma$ linearly drives the spatial hopping but inversely suppresses the effective density of states $\nu$. The cancellation of $\Gamma$ reveals that the resulting localization plateau ($\xi \sim \bar{g}_{xx}$) is robust against the change of the disorder strength.}
    \label{fig:random_walk_illust}
\end{figure*}
\subsection{Localization Length  Plateau and Universal Scaling Collapse}

In the presence of disorder ($\Gamma \neq 0$), we numerically extract the localization length $\xi$ across a wide range of disorder strengths by employing the Transfer Matrix Method (TMM) \cite{MacKinnon1981, MacKinnon1983, Kramer1993, Slevin2014}. In the conventional 1D Anderson paradigm, the localization length $\xi$ is governed by kinetic hopping. It decays monotonically with disorder strength as $\xi \propto t^2/\Gamma^2$. For benchmarking,  the localization length of a trivial band without quantum metric is shown in Fig.~\ref{fig:Lieb}c (dashed line). For a quantum geometrically trivial band with zero quantum metric, the localization length decreases rapidly towards the atomic limit once the disorder strength is comparable to the bandwidth of the band. 

However, for bands with sizable quantum metric length, the localization length reveals a clear departure from the Anderson localization behavior.
As illustrated by the solid lines in Fig.~\ref{fig:Lieb}c, in the weak-disorder regime ($\Gamma \ll 4t$), the localization length decreases rapidly with increasing disorder strength.
However, as the disorder strength exceeds the bandwidth ($\Gamma \gg 4t$), this monotonic trend breaks down. Instead of collapsing toward the atomic limit, $\xi$ saturates at a plateau which is independent of the disorder strength. This plateau persists until the disorder strength approaches the band gap $\Delta$. 
Critically, the value of this plateau is governed by the underlying quantum geometry of the band. A larger quantum metric length $l_{\mathrm{QM}}$ yields a longer localization length. This phenomenon signifies a crossover where an intrinsic length scale—rooted in the quantum metric—supersedes kinetic hopping as the dominant localizing mechanism. We term this unconventional phase \textit{quantum metric localization}, as the localization length is determined by the quantum metric length $l_{\mathrm{QM}}$.

\begin{figure*}[t!]
    \centering
    \includegraphics[width=17.2cm]{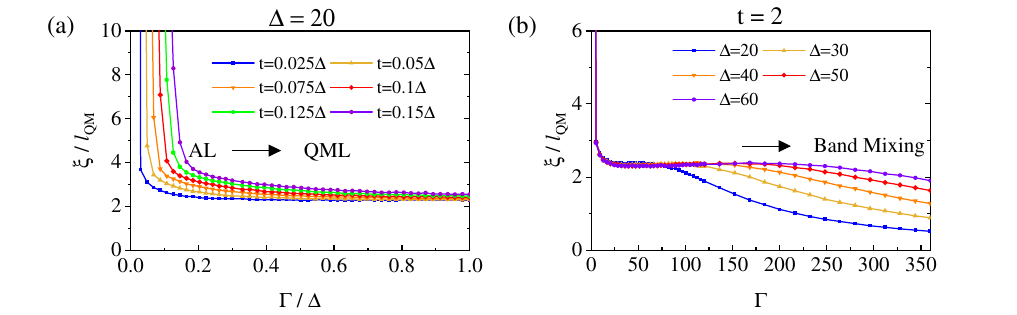}
    \caption{(Color online)
    {Quantum metric protection against disorder.} \textbf{(a)} The rescaled localization length $\xi/l_{\mathrm{QM}}$ versus disorder strength $\Gamma$ for various bandwidths of the isolated band, and the bandwidth is $4t$. AL denotes the Anderson localization, and QML denotes quantum metric localization. As disorder strength increases, there is a clear AL to QML crossover. 
    \textbf{(b)} The rescaled localization length as a function of $\Gamma$ for different values of band gap $\Delta$.  Quantum metric protection of the localization length is ineffective when $\Gamma$ is much larger than $\Delta$ when band mixing happens, such that states from faraway bands are important at the Fermi energy. Both panels utilize a fixed quantum metric length of $l_{\mathrm{QM}}=32$.} 
    \label{fig:plateau_dissolve}
\end{figure*}

Strikingly, this quantum metric localization regime exhibits a universal quantitative behavior. Upon rescaling the extracted localization lengths by $l_{\mathrm{QM}}$, as shown in Fig.~\ref{fig:Lieb}d and in Fig.~\ref{fig:plateau_dissolve}, the various plateaus obtained across different values of $l_{\mathrm{QM}}, t,$ and $\Delta$, all collapse onto a single dimensionless constant:

\begin{equation}
    \frac{\xi}{l_{\mathrm{QM}}} \sim 2.    
\end{equation}
This scaling collapse provides strong evidence that in the quantum metric localization regime, $l_{\mathrm{QM}}$ becomes the fundamental length scale governing the spatial extension of the localized wavefunctions. In previous works, we showed that the impurity induced bound state size is governed by the $l_{\mathrm{QM}}$ \cite{li2025flat,guo2025majorana,zhao2026bound}, however, showing that the localization length is governed by $l_{\mathrm{QM}}$ is indeed surprising.   

To isolate the geometric effect from possible kinetic artifacts of the Lieb lattice, we introduce a complementary two-band model in Appendix \ref{app:range2}. In this model, the quantum metric length can be continuously tuned while leaving the energy band dispersion invariant. Strikingly, this independent model exhibits a similar scaling collapse at $\xi \sim 2l_{\mathrm{QM}}$, proving that the localization plateau is governed by the quantum geometry of the isolated Hilbert space.

The formation of the localization plateau and the relation of $\xi \sim 2 l_{\mathrm{QM}}$ obtained numerically are truly surprising. The next section provides a more intuitive understanding of the localization length plateau by invoking the phenomenological Thouless formula.





\section{The Real-space Picture: The Quantum Metric Enabled Diffusion}

To understand the emergence of the localization-length plateau, we invoke Thouless formula\cite{thouless1974electrons} for the flat-band system. We begin with a semiclassical description of electron dynamics in the diffusive regime. Localization then arises as the characteristic length scale at which diffusive transport breaks down, allowing us to directly extract the localization length in one dimension from the system's diffusive properties.

Given a dispersive band with disorder, electron dynamics in the diffusive regime are governed by the diffusion coefficient scales as $D \sim v_F^2 \tau\sim v_F^2 / \Gamma$. However, in the strong disorder regime where the disorder strength overwhelms the bandwidth, the group velocity term vanishes, and the standard kinetic diffusion mechanism breaks down.
In this regime, and in the absence of interband mixing, the dynamics can be understood in a real-space picture as disorder-induced scattering between Wannier states $W(r-R)$ and $W(r-R')$, which form a complete basis for the Hilbert space of the relevant band at the Fermi energy.
The matrix element of the disorder potential between Wannier states centered at $R$ and $R'$ is given by
\begin{equation}
    V_{R,R'} = \int dr \, W^*(r-R) V(r) W(r-R').
\end{equation}
The nature of the matrix elements of the disorder potential $V(r)$ is tied to the band's quantum geometry. For a trivial quantum metric ($\overline{g}_{xx}=0$), the Wannier functions are maximally localized and non-overlapping [see Fig.~\ref{fig:random_walk_illust}(a)]. 
So the disorder matrix elements are nonzero only when $\mathbf{R} = \mathbf{R}'$, reducing the disorder to a purely diagonal random barrier potential in the Wannier basis. However, a nontrivial quantum metric introduces finite overlap between Wannier states, allowing the disorder to mediate off-diagonal transitions ($R\neq R'$) [see Fig.~\ref{fig:random_walk_illust}(b)]. In this regime, the disorder potential is no longer just a barrier but a stochastic bridge, facilitating transport through disorder-induced random hopping.
According to Fermi's Golden Rule, the ensemble-averaged transition rate is $P_{R\rightarrow R'} = \frac{2\pi}{\hslash} \overline{|V_{R,R'}|^2} \nu$. Here, the transport physics is affected by disorder in two ways. First, assuming uncorrelated Gaussian white-noise disorder, $\overline{V(r)V(r')} = \Gamma^2 \delta(r-r')$, the squared matrix element reduces to the real-space density overlap of the Wannier functions. Second, from the self-consistent Born approximation, the effective density of states $\nu$ at the Fermi level scales inversely with disorder strength: $\nu \sim 1/\Gamma$.
Combining these two contributions, the transition rate is simplified to
\begin{equation}
    P_{R\rightarrow R'} \sim \Gamma \int dr \, \rho(r-R)\rho(r-R'),
\end{equation}
where $\rho(r-R)=|W(r-R)|^2$ represents the spatial probability density of the Wannier function. 
The core of this random hopping picture lies in the  diffusion constant $D$ determined by the rate of squared displacement, summing all transition rates weighted by their squared hopping distances 
\begin{equation}
D = \sum_{R'} (R-R')^2 P_{R\rightarrow R'}\sim \Gamma \int dr \int dr' (r - r')^2 \rho(r) \rho(r'),
\label{eq:diffusion}
\end{equation}
where we transform the discrete lattice sum into a continuous spatial convolution. The integral in Eq.~\ref{eq:diffusion} turns out to be the quadratic moment of the Wannier function, $\text{Var}(W) \equiv \int dr \, \rho(r)(r - \langle r \rangle_W)^2$. 
As established by Marzari and Vanderbilt \cite{marzari1997maximally}, the quadratic moment of a maximally localized Wannier function is given by the integrated quantum metric of the Bloch states.
Therefore, we arrive at the quantum metric enabled diffusion constant
\begin{equation}
    D \sim \Gamma \bar{g}_{xx} = \Gamma l_{\mathrm{QM}}.\label{Diffusion}
\end{equation}
Therefore, disorder can activate diffusion in the presence of a quantum metric. The above semiclassical picture is consistent with wave packet dynamics and the Bethe-Salpeter equation within the ladder approximation \cite{wang2026qml}.

To obtain the localization length from the diffusion constant, we invoke the phenomenological Thouless picture for localization \cite{thouless1974electrons}. 
In one dimension, the Thouless scaling argues that the localization length $\xi$ is proportional to the product of the diffusion constant and the density of states. 
Specific to the isolated flat band system, regarding the diffusion constant $D$ in Eq.~\eqref{Diffusion} and the disorder-broadened density of states ($\nu \sim 1/\Gamma$),  the Thouless formula yields the localization length through a cancellation of the disorder strength dependence 
\begin{equation}
\xi \sim D \nu \sim (\Gamma l_{\mathrm{QM}}) \left( \frac{1}{\Gamma} \right) = l_{\mathrm{QM}}.
\end{equation}

A schematic plot of this cancellation is illustrated in Fig.~\ref{fig:random_walk_illust}c. 
The Thouless argument correctly captures the disorder independence and the scaling $\xi \propto l_{\mathrm{QM}}$, but it cannot determine the numerical prefactor. To understand the scaling behavior, the robustness of the localization plateau against disorder, and the crossover from Anderson localization to quantum metric localization, we employ a supersymmetric field theory approach in the next section.

\section{ The Supersymmetric Non-linear Sigma Model}
To understand the quantum metric localization regime analytically, we derive the macroscopic transport properties for isolated bands using the supersymmetric non-linear sigma model (SUSY NLSM) \cite{efetov1999supersymmetry}. 
We consider an isolated flat band, separated from remote dispersive bands by a large energy gap $\Delta \gg \Gamma$. The electrons are subjected to a static Gaussian disorder potential with variance $\Gamma^2$. Following standard supersymmetric techniques, we perform the Gaussian average over the disorder and integrate out the microscopic fermions [see Appendix~\ref{app:a}]. This procedure maps the disordered system onto an effective field theory described by a matrix field, $Q(q, t, t')$. This $Q$ field captures the soft diffusion modes, and its slow spatial variations govern the long-wavelength transport behavior.

We project the disorder interaction onto the isolated band. By doing so, the scattering vertices are dressed by $\Lambda(k, q)$,
\begin{equation}
     \text{str} \left( \bar{\Psi}_{k+q/2} \, Q_q  \, \Psi_{k-q/2} \right)\rightarrow  \text{str} \left( \bar{\psi}_{k+q/2} \, Q_q \Lambda(k, q)  \, \psi_{k-q/2} \right),
\end{equation}
where $\Lambda(k, q) = \langle u_{k+q/2} | u_{k-q/2} \rangle$ is the form factor of the Bloch states. Here, $\Psi_{k}$ denotes the supervector in the full Hilbert space (including all bands) at momentum $k$, whereas $\psi_{k}$ is the supervector restricted to the isolated band of interest. The effective action is formulated using the supertrace, $\text{str}(\cdot)$, with the formal construction provided in Appendix~\ref{app:a}. In the homogeneous limit $q=0$, this Bloch state overlap is simply unity. 
The quantum geometric effects emerge in the spatial gradients of the $Q$ field. 

To obtain a low-energy effective theory, we expand these spatial gradients. We assume the field $Q$ varies slowly over distances much larger than the characteristic length scale — the quantum metric length $l_{\mathrm{QM}}$ in a flat band — such that $q \cdot l_{\mathrm{QM}} \ll 1$, and that probing frequencies are much slower than the scattering rate, $\omega \ll \Gamma$.
Under the gradient expansion, the spatial stiffness is governed by the momentum expansion of the geometric form factor, $|\Lambda(k, q)|^2 \approx 1 - g_{xx}(k) q^2$, where $g_{xx}$ is the quantum metric.
By rescaling the field to a dimensionless form ($\tilde{Q}$) and keeping only the 1-loop Feynman diagrams, the effective 1D action reduces to a non-linear sigma model,
\begin{equation}
S_{NLSM}[\tilde{Q}] = -\frac{1}{2}\int dx \left[ \bar{g}_{xx} \text{str}(\partial_x \tilde{Q})^2 + \frac{2i\omega}{\Gamma} \text{str}(\Lambda_{SUSY} \tilde{Q}) \right],
\end{equation}
where $\Lambda_{SUSY} = \text{diag}(1, -1)_{RA}$ encodes the causal structure in the retarded (R) and advanced (A) sectors. 

The spatial stiffness, as the coefficient of the $\text{str}(\partial_x \tilde{Q})^2$ term, is determined by the Brillouin zone average of the quantum metric $\bar{g}_{xx} = l_{\mathrm{QM}}$. By matching our derived effective action to Efetov's canonical formulation of the 1D NLSM, $S = -\frac{L_c}{8} \int dx \, \text{str}(\partial_x \tilde{Q})^2$, we extract the macroscopic field theory parameter $L_c = 4\bar{g}_{xx}$. 
Generically, in a 1D system, the arithmetic localization length appears to be four times the typical localization length \cite{mirlin2000statistics}. 
Numerical computations via the TMM extract the Lyapunov exponent, which defines the typical wavefunction decay length, $\xi = L_c / 2$. Therefore, evaluating this projected single-band effective theory yields a plateau localization length scaled as $\xi \sim 2\bar{g}_{xx}=2l_{\mathrm{QM}}$, which is independent of the disorder strength $\Gamma$. 

We note a necessary theoretical subtlety regarding the exact quantitative prefactor predicted by this continuum field theory. As established in the literature \cite{mirlin2000statistics}, the formal validity of the NLSM gradient expansion requires a controlled intermediate diffusive window, $L_0 \ll L \ll \xi$, where $L_0$ is the microscopic mean free path. In this standard regime, the dimensionless conductance $g(L_0) \sim 2\pi \nu D L_0^{-1}$ serves as a large parameter ($g \gg 1$) that parametrically suppresses higher-loop corrections and higher-order spatial derivatives.
However, in the strong-disorder flat-band limit, diffusion is governed by the geometric overlap of adjacent Wannier functions, restricting the microscopic length scale to $L_0 \sim l_{\mathrm{QM}}$. Because the localization length is also proportional to $l_{\mathrm{QM}}$, the intermediate diffusive window is intrinsically narrow, and the dimensionless conductance evaluates to a marginal $g(L_0) \sim \mathcal{O}(1)$. In this marginally controlled regime, higher-order spatial derivatives (e.g., $\mathcal{O}(q^4)$) and microscopic lattice-scale details are not parametrically suppressed.
This limitation accounts for the quantitative discrepancies between the continuum $\sigma$-model predictions and our exact numerical lattice simulations. Nonetheless, the fundamental mathematical structure revealed by this expansion---specifically, that the dynamic cancellation of the zeroth-order terms isolates the quantum metric $g_{xx}$ as the only surviving spatial stiffness---remains robust. Thus, while the exact prefactor is subject to marginal renormalization, the field theory remains highly successful at identifying the underlying physical mechanism of the quantum metric localization plateau. This perturbative prediction is remarkably consistent with the $\xi \sim 2l_{\mathrm{QM}}$ scaling collapse observed in our exact numerical simulations.

As derived in Appendix~\ref{app:a}, introducing a dispersion $2t\cos(k)$ to the sigma model reveals a crossover between kinetic diffusion and quantum geometric spatial stiffness. In the weak disorder limit ($\Gamma \ll t$), transport is dominated by the kinetic contribution, yielding an Anderson localization. As the disorder scattering rate becomes comparable to the bandwidth ($\Gamma \sim 4t$), the kinetic contribution is quenched. As shown in Fig.~\ref{fig:plateau_dissolve}a, the localization length deviates from the Anderson prediction and collapses onto the plateau, pinned at $\xi \sim \mathcal{O}(1) l_{\mathrm{QM}}$ independent of the initial kinetic bandwidth.

This plateau is highly robust. As demonstrated numerically in Fig.~\ref{fig:plateau_dissolve}b, the plateau persists until the disorder strength exceeds the band gap. We refer to this robust survival as quantum metric protection. As shown in Appendix~\ref{app:spinlieb}, this quantum metric protection even persists against magnetic disorder in a spinful model, provided the relevant Hilbert space remains isolated.

Remote band contributions are strongly suppressed when the disorder strength is small compared to the band gap. Extending the field theory to the full multiband Hilbert space (Appendix~\ref{app:b}) demonstrates that as the disorder-dressed scattering rate (self-energy) $\tilde{\Gamma}_R$ approaches the bulk band gap ($\tilde{\Gamma}_R \sim \Delta$), the target band begins to mix with remote dispersive bands. This interband mixing effectively dilutes the geometric form factors, leading to a suppression of the effective stiffness ${g}_{eff} \propto [\Delta^2 / (\Delta^2 + \tilde{\Gamma}_R^2)]{g}_{xx}$. With the onset of severe band mixing, quantum metric protection is destroyed, and the system crosses over into atomic Anderson localization ($\xi \propto 1/\Gamma^2 $). This analytical dissolution captures the numerical breakdown shown in Fig.~\ref{fig:plateau_dissolve}b.


\section{Conclusion}
In this work, we have identified a new type of electronic localization in which the localization length is determined by the quantum metric length and is independent of disorder strength. We refer to this regime as quantum metric localization. Furthermore, the localization plateau as a function of disorder strength is exceptionally broad. The localization length does not deviate from the plateau until the disorder is so strong that states from faraway bands become important in forming the localized states at the Fermi energy. We refer to this strong protection of the localization length against disorder as quantum metric protection.

Our results show that quantum metric protection is a form of Hilbert space protection that is remarkably robust against disorder. For example, for an isolated band, the quantum metric provides a lower bound for the decay length of disorder-induced bound states when these states are formed from the Bloch states of the isolated band \cite{zhao2026bound}. 
This quantum metric thus characterizes the lower bound of the decay length, which cannot be lifted unless the perturbations are so strong that Bloch states from faraway bands become involved in forming the localized states. 
Therefore, quantum metric protection is expected to persist across diverse lattice geometries and even when certain symmetries are broken, provided the relevant Hilbert space remains isolated. This establishes quantum metric protection as a universal and robust phenomenon parallel to topological protection in gapped phases.

It is important to note that the quantum metric length can be orders of magnitude longer than the lattice spacing, depending on the physical realization and model parameters. Therefore, in physical realizations with a long quantum metric length and a sizable band gap, even strong disorder cannot localize the electrons, provided that the sample size is comparable to the quantum metric length. The discovery of quantum metric localization thus provides a new guiding principle for designing electronic, photonic, and acoustic devices.

In this work, we focus on the study of the localization problem in 1D. The interplay between quantum metric and disorders  in higher dimensions will be studied in future works.

\textit{Acknowledgements}— The authors thank Patrick Lee for insightful discussions. We also thank Zhe Hou, Chui-Zhen Chen, Pengwei Zhao, Xuzhe Ying and Yuntao Guan for discussions on the numerical calculation and field theory derivation.
   W. -B. D., J. Z., and K. T. L. acknowledge the support of the Ministry of Science and Technology, China, The New Cornerstone Foundation, the State Key Laboratory of Quantum Information Technologies and Materials, and the Hong Kong Research Grants Council through Grants No. MOST23SC01-A, No. RFS2021-6S03, No. C6053-23G, No. AoE/P-701/20, AoE/P-604/25R, No. 16309223, No. 16311424, and No. 16300325.
   
\bibliography{reference}

\section*{Appendix}
\subsection{Quantum Metric Localization in a Two-Band Model \label{app:range2}}

We introduce a two-band tight-binding model \cite{zhao2026bound} as a complement to the three-band Lieb lattice discussed in the main text. This model serves as a highly controlled platform, allowing us to isolate geometric effects from kinetic ones explicitly. By construction, we can tune the quantum metric length of the isolated band while leaving the band dispersion unchanged.

The underlying structure of the Hamiltonian is most elegantly expressed by introducing a complex momentum function $f(k) = 1 - \cos(k - \mathrm{i}\kappa)$, whose squared magnitude evaluates to $|f(k)|^2 = (\cos k - \cosh \kappa)^2$. Expanded in the Pauli basis, $\mathcal{H}(k) = d_0(k)\sigma_0 + \boldsymbol{d}(k)\cdot\boldsymbol{\sigma}$, the Hamiltonian takes a compact form
\begin{equation}
\begin{aligned}
d_0(k) &= 2t\cos(2k) + |f(k)|^2, \\
d_1(k) &= \sqrt{1-\alpha^2} \mathrm{Re}[f(k)^2], \\
d_2(k) &= \sqrt{1-\alpha^2} \mathrm{Im}[f(k)^2], \\
d_3(k) &= -\alpha |f(k)|^2.
\end{aligned}
\end{equation}




We consider the effect of the on-site Anderson disorder, represented by $\hat{V}_{x} = V(x)\sigma_0$. Here $V(x)$ is a uniformly distributed random potential with zero mean and white-noise correlation $\langle V(x)V(x')\rangle_{dis}=\Gamma^2\delta_{x,x'}$. In the clean limit ($\Gamma=0$), the eigenvalues of the matrix $\mathcal{H}(k)$ are easily deduced. The dispersion relations for the lower and upper bands are  $E_-(k) = 2t\cos (2k)$ and $E_+(k) = 2t\cos (2k) + 2|f(k)|^2$, respectively. These two bands are separated by a finite energy gap $\Delta = 2(1 - \cosh \kappa)^2$, which remains non-zero for any $\kappa\neq0$ (assuming $t\ll\Delta$). Notably, the energy spectrum is strictly invariant under changes to the parameter $\alpha$. This feature ensures that all the kinetic properties of the band structure remain fixed when tuning $\alpha$.

The parameter $\alpha$, which  does not change the energy eigenvalues, governs the underlying quantum geometric properties of the wave functions. The quantum metric $g(k)$ for the lower band is given by:
\begin{equation}
g(k) = \frac{(1-\alpha^2) \sinh^2 \kappa}{(\cos k - \cosh \kappa)^2}.
\end{equation}
Integrating this over the Brillouin zone yields the quantum metric length:
\begin{equation}\label{eq:QMLrange2}
l_{\text{QM}} = \frac{1}{2\pi} \int g(k) dk = (1-\alpha^2) \coth \kappa.
\end{equation}

By construction, $\alpha$ serves as a quantum geometric tuning parameter. This makes the model an ideal platform to isolate geometric effects from kinetic ones. In the following, we fix $\kappa$ and $t$, and vary $\alpha$ to tune $l_{\text{QM}}$. We focus on the transport properties of the lower band by setting the Fermi energy $E_F=0$ at its center. Using the transfer matrix method, we compute the localization length $\xi$ as a function of the normalized disorder strength $\Gamma/\Delta$ for various values of $l_{\text{QM}}$ with fixed kinetic bands structure.
\begin{figure*}[ht]
    \centering
\includegraphics[width=17.2cm]{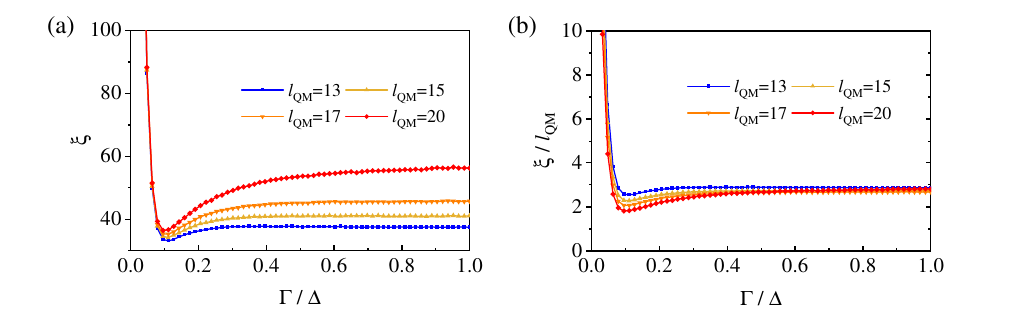}
    \caption{(Color online).   
In the two band model: localization and quantum-metric plateau with disorder.
\textbf{(a)} Disorder dependence of the localization length $\xi$, showing the emergence of a plateau.
\textbf{(b)} Rescaling by $l_{\mathrm{QM}}$ yields a collapse of $\xi/l_{\mathrm{QM}}$, demonstrating a parameter-independent localization plateau across different quantum metric lengths. Numerical parameters: $\kappa = 0.04$, $t=0.1\Delta=0.2(1-\cosh\kappa)^2$ with $l_{\text{QM}}$ tuned via $\alpha$ through Eq.~\ref{eq:QMLrange2}.
}
    \label{fig:range2}
\end{figure*}

As shown in Fig.~\ref{fig:range2}a, in the weak-disorder regime, $\xi$ follows the Anderson scaling governed by kinetic processes. However, as $\Gamma$ increases and overcomes the residual bandwidth, $\xi$ deviates from the Anderson prediction and develops a plateau. Even though the bands structures remain strictly identical for all curves, the magnitude of the localization plateau shifts significantly as $l_{\text{QM}}$ varies. 
The quantum geometric origin of this behavior is further elucidated in Fig.~\ref{fig:range2}b. By rescaling the localization lengths with the quantum metric length $l_{\text{QM}}$, we find that those curves for different $l_{\text{QM}}$ values approximatly collapse onto a single plateau with $\xi / l_{\text{QM}} \sim \mathcal{O}(1)$. 

\subsection{Spinful Lieb lattice with magnetic Disorder\label{app:spinlieb}}
\begin{figure*}[ht]
    \centering
\includegraphics[width=17.2cm]{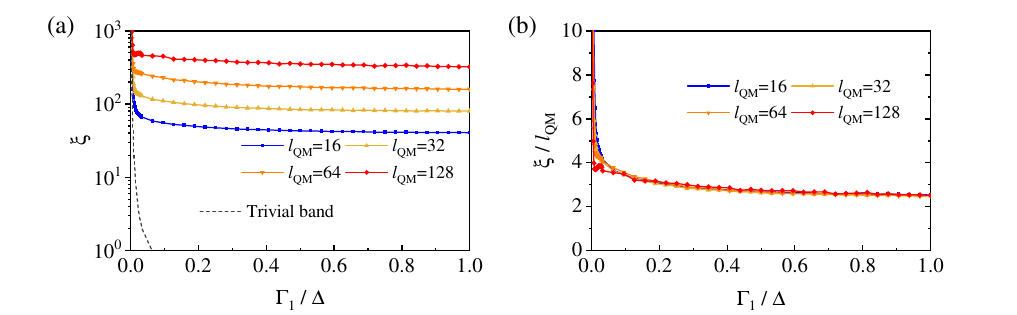}
    \caption{(Color online).   
In the spinful 1D Lieb lattice: localization and quantum-metric plateau with  magnetic disorder.
\textbf{(a)}  Dependence of the localization length $\xi$ on disorder strength. In the regime where disorder exceeds the central bandwidth parameter $2\lambda$, trivial bands cross over into an atomic insulating limit. 
\textbf{(b)} The ratio $\xi/l_{\mathrm{QM}}$. 
Rescaling by $l_{\mathrm{QM}}$ yields a collapse of $\xi/l_{\mathrm{QM}}$. Numerical parameters: $t=0$, $\Delta = 20$ and $\lambda=0.03\Delta$ with $l_{\text{QM}}$ tuned via $\delta = 1 + 4l_{\mathrm{QM}} - 2\sqrt{2l_{\mathrm{QM}}(1 + 2l_{\mathrm{QM}})}$ and $J = \Delta/(2\sqrt{2}\delta)$.}
    \label{fig:spinLieb}
\end{figure*}

For a third example, we further consisder a spinful model with spin-orbtial coupling with magnetic disorder. The Hamiltonian is given by
\begin{equation}
\begin{split}
H = H_{\text{Lieb}} \sigma_0 & + \lambda \sum_k \mathbf{a}_k^\dagger \left( \sin k \, \sigma_3 \right) \mathbf{a}_k \\
& + \sum_x \mathbf{c}_x^\dagger V_1(x) \mathbf{I}_{3 \times 3} \sigma_1 \mathbf{c}_x,
\end{split}
\end{equation}
where $H_{\text{Lieb}}$ is defined as Eq.~\ref{eq:Hlieb} of the main text in the clean limit and $\mathbf{a}_{k}$ denotes the basis in momentum space. 
Here $\sigma_{1,2,3}$ denote the Pauli matrices acting in the spin-$1/2$ space, and $\sigma_0$ is the identity. The spin-orbit coupling term $\sin(k) \sigma_3$ introduces momentum-dependent spin splitting. $V_{1}(x)$ is the random potential for the $\sigma_1$-type magnetic disorder, following a uniform distribution with zero mean and white-noise correlation $\langle V_1(x)V_1(x') \rangle_{\text{dis}} = \Gamma_1^2 \delta_{x,x'}$, where $\Gamma_1$ denotes the disorder strength.

There are two bands near the fermi energy $E_F=0$ with a spin-dependent dispersion $E_{\uparrow,\downarrow}(k) = \pm \lambda \sin k$. They remain separated from the other bands by a finite energy gap of order $\Delta + \mathcal{O}(\lambda)$. Despite the multiband nature of the problem, the quantum geometry of the relevant subspace can still be characterized via the band-resolved quantum metric, from which we define the quantum metric length $l_{\mathrm{QM}}$ through Brillouin zone averaging, as in Eq.~\ref{eq:lqm}.

Using the transfer matrix method, we compute the localization length $\xi$ as a function of $\Gamma_1/\Delta$ in Fig.~\ref{fig:spinLieb}a and b. 
As shown in Fig.~\ref{fig:spinLieb}a, in the weak-disorder regime, $\xi$ follows Anderson scaling governed by kinetic processes. However, when $\Gamma_1$ is much larger than the characteristic bandwidth $2\lambda$, the localization length deviates from the Anderson prediction and develops a disorder-independent plateau, which persists until $\Gamma_1$ exceeds the band gap $\Delta$.
As shown in Fig.~\ref{fig:spinLieb}, the localization lengths for various parameters collapse onto a value $\xi / l_{\mathrm{QM}} \sim \mathcal{O}(1)$.
This data collapse is consistent with observations in the Lieb lattice (Fig.~\ref{fig:Lieb}d), demonstrating that the localization plateau and its scaling are robust against spin-orbit coupling and magnetic disorder, and persist even in complex multiband systems.

\subsection{Localization of finite dispersion in the weak disorder limit\label{app:a}}

We consider a 1D system preserving time-reversal symmetry (TRS), with a narrow isolated band having kinetic dispersion $\epsilon_0(k) = 2t \cos(k)$, separated from other bands by a large energy gap $\Delta \gg t$. To compute disorder-averaged observables without replica indices, we employ the supersymmetric (SUSY) method \cite{efetov1999supersymmetry}. We introduce an 8-component supervector $\Psi_{k}(t)$ containing commuting (bosonic) and anti-commuting (fermionic) variables, expanded to encompass the retarded and advanced causal sectors as well as the time-reversal (Nambu) sector to simultaneously capture both diffuson and cooperon modes. In the momentum basis, the clean partition function is represented by the non-interacting action
\begin{equation}
S_0 = \int dt \sum_{k} \bar{\Psi}_{k} G_0^{-1} \Psi_{k},
\end{equation}
where $G_0^{-1} = i\partial_t - \epsilon(k)$ is the bare inverse Green's function. The system is subjected to a quenched, uncorrelated Gaussian scalar disorder potential $V_q$ with zero mean and variance $\langle V_q V_{-q} \rangle = \Gamma^2$. The disorder action takes the form
\begin{equation}
S_{dis} = -\sum_{k,q} \int dt \, \bar{\Psi}_{k+q/2}(t) V_q \Psi_{k-q/2}(t).
\end{equation}
To perform the exact disorder average of the generating functional, we trace out the Gaussian distribution of $V_q$. Completing the square in the functional path integral mathematically generates an effective four-superfermion interaction. Because the quenched disorder is static, it perfectly correlates scattering events across independent times $t$ and $t'$,
\begin{widetext}
\begin{equation}
\langle e^{iS_{dis}} \rangle _{dis}= \exp\left[ -\frac{\Gamma^2}{2} \int dt dt' \sum_q \left( \sum_k \bar{\Psi}_{k+q/2}(t) \Psi_{k-q/2}(t) \right) \left( \sum_{k'} \bar{\Psi}_{k'-q/2}(t') \Psi_{k'+q/2}(t') \right) \right].
\end{equation}
\end{widetext}

Applying the Fierz super-algebraic rearrangement pairs the times $t$ and $t'$, allowing us to decouple this quartic interaction via a Hubbard-Stratonovich transformation using an auxiliary supermatrix field $Q(q, t, t')$. In this decoupling scheme, $Q_q$ carries the center-of-mass momentum transfer $q$ and acts as a conjugate variable to the total local density, effectively summing over the internal Bloch momentum $k$. The decoupling action is
\begin{equation}
S_{HS} = \int \mathcal{D}[Q] \exp \left[ -\frac{1}{2\Gamma^2} \text{str}(Q^2) - i \text{str} \left( \bar{\Psi} Q \Psi \right) \right],
\end{equation}
where $\text{str}$ denotes the supertrace, and the contraction $\bar{\Psi} Q \Psi$ implicitly contains the double time integral $\int dt dt' \, \bar{\Psi}(t) Q(t,t') \Psi(t')$. By trading the static microscopic field $V_q$ for the dynamic supermatrix field $Q$, we shift to the hydrodynamic degrees of freedom. While the $\delta$-correlated disorder is rapidly oscillating, the $Q$-matrix varies slowly in space and time, effectively capturing the propagating diffusive modes. Assuming all other bands are sufficiently remote ($\Delta \gg \max\{\Gamma, t\}$), we construct the effective theory purely within the target Hilbert space. We project onto the Bloch band by forcing
\begin{equation}
\Psi_k \rightarrow \ket{u_{k}} \psi_k, \quad \text{and} \quad \bar{\Psi}_k \rightarrow \bar{\psi}_k \bra{u_{k}},
\end{equation}
where $\ket{u_{k}}$ is the Bloch state of the target band. Substituting this projection into the action, the scattering vertex is naturally dressed by the form factor $\Lambda(k, q) \equiv \langle u_{k+q/2} | u_{k-q/2} \rangle$,
\begin{equation}
S_{coupling} = -i \sum_{k,q} \text{str} \left( \bar{\psi}_{k+q/2} \, Q_q \Lambda(k, q) \, \psi_{k-q/2} \right).
\end{equation}
The fermions now only appear quadratically. We exactly integrate them out using the rules of Gaussian super-integration, yielding the exact effective action for the $Q$-matrix,
\begin{equation} \mathcal{F}[Q] = \frac{1}{2\Gamma^2 } \text{str}(Q^2) + \text{str} \ln \left[ G_0^{-1}- Q \Lambda \right] .\end{equation}

To identify the stationary state, we extremize the free energy functional. The saddle point $Q_0$ represents the homogeneous, static mean-field ground state. In this uniform limit ($q=0$), the geometric form factor strictly evaluates to unity, $\Lambda(k, 0) = \langle u_k | u_k \rangle = 1$, demonstrating that the mean-field scattering rate is a local property independent of the underlying quantum geometry. The variation $\delta F / \delta Q = 0$ yields the exact saddle-point equation
\begin{equation}
\frac{Q_0}{\Gamma^2} = \sum_k \frac{1}{-\epsilon_0(k) - Q_0}.
\end{equation}
Because static disorder does not mix causal sectors on average, the saddle point is block-diagonal: $Q_0 = \text{diag}(\Sigma^R, \Sigma^A)$. To preserve causality (poles residing in the appropriate complex half-planes), the saddle point is locked into the matrix structure $Q_0 = -i \tilde{\Gamma}_R \Lambda_{SUSY}$, where $\Lambda_{SUSY} = \text{diag}(1, -1)_{RA}$. The momentum summation over the 1D Brillouin zone becomes an exact contour integral
\begin{equation} 
\tilde{\Gamma}_R = i\Gamma^2 \int_{-\pi}^{\pi} \frac{dk}{2\pi} \frac{1}{-2t\cos(k) + i\tilde{\Gamma}_R} = \frac{\Gamma^2}{\sqrt{\tilde{\Gamma}_R^2 + 4t^2}} .
\end{equation}
Squaring both sides and solving for the exact disorder-dressed energy scale $\tilde{\Gamma}_R$ yields
\begin{equation} 
\tilde{\Gamma}_R = \sqrt{\sqrt{4t^4 + \Gamma^4} - 2t^2} .
\end{equation}

To capture transport, we consider fluctuations around the homogeneous saddle point.
Longitudinal fluctuations alter the eigenvalues of $Q$ (violating local probability conservation) and acquire a massive energy penalty proportional to $\tilde{\Gamma}_R$. In the limit when probing length scales much larger than the quantum metric length scale ($L \gg \bar{g}_{xx}$) 
and frequencies much smaller than the scattering rate ($\omega \ll \tilde{\Gamma}_R$), these massive modes are integrated out. The remaining soft modes are transverse, continuously rotating the saddle point while strictly obeying the non-linear constraint $Q^2 = -\tilde{\Gamma}_R^2$. This geometric constraint mathematically enforces the Ward Identity.

The physical fluctuations are parameterized by $Q(r, t) = T^{-1} Q_0 T = Q_0 \exp(W(r, t))$. The theory preserves the full orthosymplectic global symmetry $G = \text{OSp}(2,2|4)$, while $Q_0$ is invariant under the causality-preserving maximal subgroup $H = \text{OSp}(2|2) \times \text{OSp}(2|2)$. Consequently, the transverse Goldstone modes fluctuate entirely within the off-diagonal coset manifold $\mathcal{M} = G/H$.

We perform a gradient expansion of the trace logarithm in the slow spatial variations $q$ and external probing frequencies $\omega$. Expanding the temporal sector to leading order for small $\omega$ yields $- \text{str}(\omega \Lambda_{SUSY} (\epsilon_0(k) + Q)^{-1})$. Utilizing the exact non-linear saddle-point identity $\sum_k(\epsilon_0(k) + Q)^{-1} = -Q / \Gamma^2$, the temporal action evaluates to
\begin{equation}
S_\omega[Q] = \frac{\omega}{\Gamma^2 } \int dx \, \text{str}(\Lambda_{SUSY} Q).
\end{equation}

For the spatial sector, the conventional semiclassical gradient expansion (relying on $\mathbf{v}_F$) breaks down because $\Gamma \sim t$. Instead, spatial gradients emerge simultaneously from two independent microscopic origins: the residual kinetic group velocity ($\nabla_k \epsilon_0(k)$) and the geometric spatial overlap of the Wannier orbitals ($\Lambda(k, q)$).
Isolating the finite-momentum fluctuations via $Q_{total}(q) = Q_0\delta_{q,0} +\delta Q_q$, we factor out the disorder-dressed Green's function $G_k = (G_0^{-1} - Q_0)^{-1}$,
\begin{equation}
S = \text{str} \ln \left[ G_k^{-1} - \delta Q_q \Lambda(k, q) \right].
\end{equation}
The zero-order term $O(q^0)$ vanishes by the saddle-point condition. The first-order spatial term $O(q^1)$ is odd in $k$ due to inversion and time-reversal symmetries, vanishing identically upon Brillouin zone integration. The leading transport physics emerges precisely at second order
\begin{equation}
S_2 = -\frac{1}{2} \sum_{k, q} \text{str} \left[ G_{k+q/2} \, \delta Q_q \Lambda(k, q) \, G_{k-q/2} \, \delta Q_{-q} \Lambda(k, -q) \right].
\end{equation}
Because $\delta Q_q$ anti-commutes with $Q_0$, it forces a causal block-flip, naturally isolating the disorder-dressed polarization bubble
\begin{equation}
S_2 = -\frac{1}{2} \sum_q \text{str}\left[ \delta Q_q \delta Q_{-q} \right] \sum_k |\Lambda(k, q)|^2 G^R_{k+q/2} G^A_{k-q/2}.
\end{equation}
Evaluating $S_2$ at $\mathcal{O}(q^2)$ and $\omega=0$, the geometric form factor expands via the quantum metric
\begin{equation}
|\Lambda(k, q)|^2 \approx 1 - g_{xx}(k) q^2 + \mathcal{O}(q^4).
\end{equation}
Simultaneously, we evaluate the paired causal Green's functions $G^{R/A}_k = (\pm i\tilde{\Gamma}_R - 2t \cos(k))^{-1}$ by transforming the momentum convolution into a contour integral over the unit circle $z = e^{ik}$
\begin{widetext}
\begin{equation}
    B(q)=\sum_k G^R_{k+q/2} G^A_{k-q/2} =\frac{1}{t^2} \oint_{|z|=1} \frac{dz}{2\pi i} \frac{z}{\left(z^2 e^{iq} - i\frac{\tilde{\Gamma}_R}{t} z e^{iq/2} + 1\right)\left(z^2 e^{-iq} + i\frac{\tilde{\Gamma}_R}{t} z e^{-iq/2} + 1\right)}.
\end{equation}
\end{widetext}
Applying the residue theorem yields
\begin{equation}
B(q) = \frac{\tilde{\Gamma}_R^2}{\Gamma^2 \left[ \tilde{\Gamma}_R^2 + 4t^2 \sin^2(q/2) \right]}\approx \frac{1}{\Gamma^2}\left(1-\frac{t^2}{\tilde{\Gamma}_R^2}q^2\right)+\mathcal{O}(q^4).
\end{equation}
The unified polarization bubble $\Pi(q)$ is achieved by multiplying the geometric and kinetic expansions to $\mathcal{O}(q^2)$. 
\begin{equation}
\Pi(q) \approx \frac{1}{\Gamma^2} - q^2 \left[ \sum_k \frac{g_{xx}(k)}{\epsilon_0(k)^2 + \tilde{\Gamma}_R^2} + \frac{t^2}{\Gamma^2 \tilde{\Gamma}_R^2} \right] + \mathcal{O}(q^4).
\end{equation}
The zeroth-order term strictly evaluates to $1/\Gamma^2$ by the exact saddle-point equation. The $\mathcal{O}(q^2)$ term identifies the exact unified spatial stiffness of the system, comprising both the geometrically-weighted and exact kinetic contributions.

Regarding whether higher-order expansions of the trace logarithm, $\mathcal{O}(\delta Q^n)$ for $n > 2$, generate independent $\mathcal{O}(q^2)$ spatial operators that fail to assemble into the canonical diffusion gradient, we guarantee the formal absence of such terms by the global continuous symmetry of the Goldstone target manifold $\mathcal{M} = G/H$. The field $Q = T^{-1} Q_0 T = Q_0 \exp(W)$ represents a continuous local rotation of the degenerate vacuum. By Goldstone's theorem, the action vanishes for spatially uniform rotations ($q=0$), ensuring that the effective action depends on spatial derivatives. Furthermore, any generated operator must remain invariant under the global transformations of the full supergroup $G$. On a symmetric coset space, the unique geometric scalar invariant containing exactly two spatial derivatives is the metric contraction: $\text{str}(\nabla Q)^2$. Consequently, there are no other mathematically permissible $\mathcal{O}(q^2)$ operators. When the trace logarithm is expanded to higher orders in $\delta Q$, the ensuing $q^2$-dependent terms (representing higher-order multi-impurity scattering diagrams) do not form new gradient operators. Instead, by the Ward identities, they perfectly reconstruct the higher-order non-linear terms of the target manifold's Taylor expansion, $\text{str}(\nabla Q)^2 \sim \text{str}(\nabla W)^2 + \mathcal{O}(W^2 (\nabla W)^2)$. Therefore, the spatial stiffness coefficient extracted from the second-order polarization bubble defines the prefactor for the exact non-linear gradient action, directly yielding the diffusion operator $\text{str}(\nabla Q)^2$. While higher-order expansions of the trace logarithm and the geometric form factor $\Lambda(k,q)$ generate higher-derivative terms such as $\mathcal{O}(q^4) \sim \text{str}(\partial^2_x Q)^2$, standard Renormalization Group (RG) analysis shows that these higher-gradient operators are irrelevant. They govern short-distance microscopic details but scale to zero in the long-wavelength limit. Truncating the effective action at the $\mathcal{O}(q^2)$ polarization bubble is sufficient to capture the asymptotic transport of localization physics. 

Now we can map to a dimensionless supermanifold where the target space is strictly $\tilde{Q}^2 = \mathbb{I}$. We use the exact saddle point to define the substitution $Q(x) = -i \tilde{\Gamma}_R \tilde{Q}(x)$. The complete non-linear sigma model action for the dispersive band is
\begin{widetext}
\begin{equation} 
S_{NLSM}[\tilde{Q}] = -\frac{\tilde{\Gamma}_R}{2\Gamma^2}\int dx \left[ \left( \sum_k \frac{\Gamma^2\tilde{\Gamma}_Rg_{xx}(k)}{\epsilon_0(k)^2 + \tilde{\Gamma}_R^2} + \frac{t^2}{\tilde{\Gamma}_R} \right) \text{str}(\nabla \tilde{Q})^2 + 2i\omega \text{str}(\Lambda_{SUSY} \tilde{Q}) \right] .
\end{equation}
\end{widetext}

By matching the coefficients, we explicitly extract the diffusion tensor
\begin{equation} 
D = \sum_k \frac{\Gamma^2\tilde{\Gamma}_Rg_{xx}(k)}{\epsilon_0(k)^2 + \tilde{\Gamma}_R^2} + \frac{t^2}{\tilde{\Gamma}_R}  .
\end{equation}
In one spatial dimension, the functional path integral of a statistical field theory maps exactly onto the Schrödinger equation of a quantum mechanical particle moving on the curved target manifold $\mathcal{M} = G/H$. Utilizing Efetov's exact solution on the highly symmetric $\text{OSp}(2,2|4)$ coset space \cite{efetov1983kinetics}, the localization length $L_c$ for a canonically defined 1D-NLSM action $S = -\frac{L_c}{8} \int dx \, \text{str}(\nabla \tilde{Q})^2$ can be extracted analytically. Matching our derived unified stiffness to this canonical form, the exact 1D localization length is
\begin{equation} 
L_c = 4 \left( \sum_k g_{xx}(k) \frac{\tilde{\Gamma}_R^2}{\epsilon_0(k)^2 + \tilde{\Gamma}_R^2} + \frac{t^2}{\Gamma^2} \right).
\end{equation}
This expression captures both asymptotic regimes. When the disorder strength $\Gamma$ far exceeds the characteristic bandwidth ($\Gamma\gg t$), The Lorentzian spectral weight flattens across the Brillouin zone, $\tilde{\Gamma}_R^2 / (\epsilon_0(k)^2 + \tilde{\Gamma}_R^2) \to 1$ and the kinetic channel vanishes, recovering the purely geometry-driven localization of the exact flat band: $L_c = 4\bar{g}_{xx}$. 

According to standard scaling theory\cite{mirlin2000statistics}, the rigorous validity of the continuum NLSM relies on the existence of a parametrically broad intermediate diffusive window, $L_0 \ll L \ll \xi$, where $L_0$ represents the microscopic UV cutoff. This scale separation is mathematically guaranteed if the bare dimensionless conductance at the matching scale is large, $g(L_0) \sim 2\pi \nu D L_0^{-1} \gg 1$, which justifies truncating the field theory's gradient expansion at second order ($\mathcal{O}(q^2)$). However, in an isolated 1D flat band, the diffusion constant scales as $D \propto \bar{g}_{xx} \Gamma$ and the density of states as $\nu \propto 1/\Gamma$. Numerically, it shows $L_0 \sim l_{\mathrm{QM}}$, the explicit disorder dependence perfectly cancels out, yielding a bare conductance $g(L_0) \sim \mathcal{O}(1)$. Because the expansion parameter $1/g(L_0)$ is not small, the nominal diffusion window is narrow, and the continuum theory is marginally controlled. In this regime, higher-order spatial derivatives (e.g., $\mathcal{O}(q^4)$) and microscopic lattice details are not parametrically suppressed. This limitation naturally accounts for the quantitative discrepancies—such as the differing $\mathcal{O}(1)$ prefactors—between the continuum $\sigma$-model predictions and our exact numerical results, while the field theory remains highly successful at capturing the qualitative crossover physics.

\subsection{Cancellation due to dispersive bands in the strong disorder limit}\label{app:b}

We consider a two-band system characterized by the bare Green's function in the band basis, ${G}_0^{-1} = i\omega - {\epsilon}$, where ${\epsilon} = \text{diag}(0, \epsilon(k))$ describes a strictly flat band and a remote dispersive band. Following the standard SUSY disorder averaging and Hubbard-Stratonovich decoupling, we arrive at the analogous effective free energy
\begin{equation} 
F[Q] = \frac{1}{2\Gamma^2 } \text{str}(Q^2) + \text{str} \ln \left[ {G}_0^{-1}- Q {\Lambda} \right] .
\end{equation}
In this extended Hilbert space, ${\Lambda}$ becomes a matrix tracking both intra-band and inter-band scattering
\begin{equation}
{\Lambda} = \begin{pmatrix} \Lambda_{ff} & \Lambda_{fd} \\\Lambda_{df} & \Lambda_{dd} \end{pmatrix},
\end{equation}
where $\Lambda_{nm}(k, q) = \langle u^n_{k+q/2} | u^m_{k-q/2} \rangle$ are the geometric form factors, with the indices $n,m \in \{f,d\}$ denoting the flat ($f$) and dispersive ($d$) bands, respectively.

At the homogeneous saddle point (${q}=0, \omega=0$), the Bloch states are orthogonal, rendering the form factor purely diagonal: $\Lambda_{nm}(0) = \delta_{nm}$. The mean-field saddle-point equation thus decouples into a sum over the band subspace
\begin{equation}
\frac{Q_0}{\Gamma^2} = \frac{1}{-Q_0} + \sum_k \frac{1}{-\epsilon(k) - Q_0}.
\end{equation}
To preserve the causal symmetries of the field theory, the saddle point must take the form $Q_0 = -i\tilde{\Gamma}_R \Lambda_{SUSY}$. Substituting this ansatz and isolating the $\Lambda_{SUSY}$ components yields the self-consistency equation for the disorder-dressed scattering rate $\tilde{\Gamma}_R$
\begin{equation}
\frac{\tilde{\Gamma}_R}{\Gamma^2} = \frac{1}{\tilde{\Gamma}_R} + \sum_k\frac{\tilde{\Gamma}_R}{\epsilon(k)^2 + \tilde{\Gamma}_R^2}.
\end{equation}
We evaluate the asymptotic limits of this self-consistency equation to understand the dressed energy scale:
\begin{itemize}
    \item Let $\Delta \equiv \min_{k}|\epsilon(k)|$ define the energy gap between the flat band and the dispersive band. In the weak disorder limit ($\Gamma \ll \Delta$), the inter-band mixing is negligible, and the equation exactly recovers the isolated flat-band self-energy: $\tilde{\Gamma}_R \approx \Gamma$.
    \item Let $W_d \equiv \max_k{| \epsilon(k)|}$ represent the bandwidth of the whole system. In the strong disorder limit ($\Gamma \gg W_d$), the scattering rate dominates over the kinetic dispersion, yielding $\tilde{\Gamma}_R \approx \sqrt{2}\Gamma$.
\end{itemize}
Evaluated at the saddle point, the disorder-dressed Green's functions for the respective bands are
\begin{equation}
G_f = (-Q_0)^{-1} = \frac{Q_0}{\tilde{\Gamma}_R^2},
\end{equation}
\begin{equation}
G_d = (-\epsilon(k)- Q_0)^{-1} = \frac{-\epsilon(k) + Q_0}{\epsilon(k)^2 + \tilde{\Gamma}_R^2}.
\end{equation}

To extract the spatial stiffness, we expand the trace logarithm to second order in the spatial momentum $q$. The quadratic term generates the polarization bubble, which now explicitly incorporates interband mixing
\begin{equation}
-\frac{1}{2}\text{str}\left(G_n \delta Q_q G_m \delta Q_{-q} |\Lambda_{nm}(q)|^2 \right).
\end{equation}
We first isolate the geometric contributions by expanding the form factors $|\Lambda_{nm}(q)|^2$. Because the two bands form a completely isolated subspace, the conservation of total quantum geometry implies that the inter-band Berry connections must exactly sum to the intra-band quantum metric. Letting ${g}_{xx}(k)$ denote the quantum metric of the flat band, the small-$q$ expansions are:
\begin{itemize}
    \item $|\Lambda_{ff}|^2 \approx 1 - {g}_{xx}(k) q^2$,
    \item $|\Lambda_{dd}|^2 \approx 1 - {g}_{xx}(k) q^2$,
    \item $|\Lambda_{fd}|^2 = |\Lambda_{df}|^2 \approx +{g}_{xx}(k) q^2$.
\end{itemize}
By substituting the dressed Green's functions $G_f$ and $G_d$, the causal supertrace $\text{str}(G_n \delta Q_q G_m \delta Q_{-q})$ yields the dynamic coefficients:
\begin{itemize}
    \item $C_{ff} = \frac{1}{\tilde{\Gamma}_R^2}$,
    \item $C_{dd} = \frac{1}{\epsilon(k)^2 + \tilde{\Gamma}_R^2}$,
    \item $C_{fd} = C_{df} = \frac{1}{\epsilon(k)^2 + \tilde{\Gamma}_R^2}$.
\end{itemize}

Summing these coupled contributions, $K = -\frac{1}{2}\sum_{n,m} C_{nm} |\Lambda_{nm}|^2$, the zeroth-order ($q^0$) mass terms dynamically cancel due to the non-linear Goldstone constraint, isolating the geometric spatial stiffness
\begin{equation}
K^{geo} =\sum_k\frac{1}{2}{g}_{xx}(k) \left[ \frac{\epsilon(k)^2}{\tilde{\Gamma}_R^2\left(\epsilon(k)^2 + \tilde{\Gamma}_R^2\right)} \right].
\end{equation}

Next, we evaluate the kinetic stiffness arising from the dispersive band by expanding the bare dispersion: $\epsilon({k\pm q/2})
=
\epsilon(k)
\pm \frac{q}{2}v_k
+
\frac{q^2}{8}a_k
+\mathcal O(q^3)$, where $v_k \equiv \partial_k \epsilon(k)$ is the group velocity, $a_k\equiv \partial_k^2\epsilon(k)$ is the second derivative. The cross-band kinetic contributions ($\Pi_{fd}$ and $\Pi_{df}$) vanish at $\mathcal{O}(q^2)$ because the inter-band form factor vanishes at the saddle point $\Lambda_{df}(0)=0$. 
The kinetic contribution can be obtained by expanding the full causal bubble
\begin{equation}
\begin{split}
\Pi_{dd}(q)&=\sum_k G^R_{k+q/2}G^A_{k-q/2}\\
&=\sum_k\frac{1}{-\epsilon_{k+q/2}+i\tilde{\Gamma}_R}\cdot\frac{1}{-\epsilon_{k-q/2}-i\tilde{\Gamma}_R}.
\end{split}
\end{equation}
Substituting the expansion of the dispersion, one obtains
\begin{widetext}    
\begin{equation}
\begin{split}
\Pi_{dd}(q)
=
\frac{1}{(\epsilon(k)^2+\tilde{\Gamma}_R^2)}
+
\frac{i\tilde{\Gamma}_R v_k}{(\epsilon(k)^2+\tilde{\Gamma}_R^2)^2}q
+
q^2
\left[
\frac{v_k^2(\epsilon(k)^2-3\tilde{\Gamma}_R^2)}
{4(\epsilon(k)^2+\tilde{\Gamma}_R^2)^3}
-
\frac{\epsilon(k) a_k}{4(\epsilon(k)^2+\tilde{\Gamma}_R^2)^2}
\right]
+
\mathcal O(q^3).
\end{split}
\end{equation}
\end{widetext}
The term linear in \(q\) vanishes after Brillouin-zone integration under inversion/time-reversal symmetry. 
Using the periodicity of the Brillouin zone, the second-order term may be integrated by parts:
\begin{equation}
\sum_k
\frac{\epsilon(k)\partial_k^2 \epsilon(k)}
{4(\epsilon(k)^2+\tilde{\Gamma}_R^2)^2}
=
\sum_k
\left[
\frac{\epsilon(k)^2v_k^2}
{(\epsilon(k)^2+\tilde{\Gamma}_R^2)^3}
-
\frac{v_k^2}
{4(\epsilon(k)^2+\tilde{\Gamma}_R^2)^2}
\right].
\end{equation}
Substituting this identity gives 
\[
\Pi_{dd}(q)
=
\Pi_{dd}(0)
-\frac{q^2}{2}
\sum_k
\frac{v_k^2}
{\left(\epsilon(k)^2+\tilde{\Gamma}_R^2\right)^2}
+
\mathcal O(q^4).
\]
The kinetic stiffness is therefore
\[
K^{\rm kin}
=
\frac14
\sum_k
\frac{v_k^2}
{\left(\epsilon(k)^2+\tilde{\Gamma}_R^2\right)^2}.
\]

Rescaling the field to the canonical dimensionless manifold, $\tilde{Q} = iQ / \tilde{\Gamma}_R$, we assemble the exact NLSM action for the two-band system
\begin{widetext}
\begin{equation} 
S_{NLSM}[\tilde{Q}] = -\frac{\tilde{\Gamma}_R}{2\Gamma^2}\int dx \left[ \sum_k\left[ {g}_{xx}(k) \frac{\Gamma^2\epsilon(k)^2}{\tilde{\Gamma}_R\left(\epsilon(k)^2 + \tilde{\Gamma}_R^2\right)} +\frac{v_k^2\Gamma^2\tilde{\Gamma}_R}
{2\left(\epsilon(k)^2+\tilde{\Gamma}_R^2\right)^2} \right] \text{str}(\nabla \tilde{Q})^2 + 2i\omega \text{str}(\Lambda_{SUSY} \tilde{Q}) \right].
\end{equation}
\end{widetext}

By matching the prefactors, we explicitly extract the diffusion tensor:
\begin{equation} 
D =\sum_k\left[ {g}_{xx}(k) \frac{\Gamma^2\epsilon(k)^2}{\tilde{\Gamma}_R\left(\epsilon(k)^2 + \tilde{\Gamma}_R^2\right)} +\frac{v_k^2\Gamma^2\tilde{\Gamma}_R}
{2\left(\epsilon(k)^2+\tilde{\Gamma}_R^2\right)^2} \right].
\end{equation}

Applying Efetov's exact zero-dimensional transfer operator solution for the 1D SUSY NLSM, the exact 1D localization length evaluates to
\begin{equation}
    L_c = 4 \left( \sum_k\left[ {g}_{xx}(k) \frac{\epsilon(k)^2}{\left(\epsilon(k)^2 + \tilde{\Gamma}_R^2\right)} +\frac{v_k^2\tilde{\Gamma}_R^2}
{2\left(\epsilon(k)^2+\tilde{\Gamma}_R^2\right)^2}  \right]\right).
\end{equation} 

This analytical solution captures the two transport limits of the multiband system:
\begin{itemize}
    \item \textbf{Isolated Flat Band ($\Gamma \ll \Delta$):} The disorder is insufficient to bridge the band gap ($\tilde{\Gamma}_R \approx \Gamma$). The kinetic contributions from the remote band vanish, and the interband factor $\epsilon^2/(\epsilon^2 + \tilde{\Gamma}_R^2) \to 1$. We seamlessly recover the geometry-driven localization of the isolated flat band: $L_c \rightarrow 4\bar{g}_{xx}$.
    \item \textbf{Interband Mixing ($\Gamma \gg W_d$):} The strong disorder scattering completely washes out the band structure. The geometric stiffness is suppressed by a factor of $\sim W_d^2/\Gamma^2$, signaling a collapse of the quantum metric protection. The localization length vanishes as in a normal unprotected band: $L_c \sim \frac{1}{\Gamma^2}\rightarrow0$.
\end{itemize}

\end{document}